\documentclass[nobibnotes,preprintnumbers,aps,prl,superscriptaddress,showpacs,twocolumn,twoside,sort&compress]{revtex4}				 %
\usepackage{graphicx}																												 %
\usepackage{titlesec}																											     %
\usepackage{fancyheadings}																							 				 %
\begin{document}																													 %
\preprint{{Vol.XXX (201X) ~~~~~~~~~~~~~~~~~~~~~~~~~~~~~~~~~~~~~~~~~~~~~~~~~~~~ {\it CSMAG`16}										  ~~~~~~~~~~~~~~~~~~~~~~~~~~~~~~~~~~~~~~~~~~~~~~~~~~~~~~~~~~~~ No.X~~~~}}																 %
\vspace*{-0.3cm}																													 %
\preprint{\rule{\textwidth}{0.5pt}}																											 \vspace*{0.3cm}																														 %

\title{Theoretical Study of the Frustrated Ising Antiferromagnet on the Honeycomb Lattice  }

\author{A. Bob\'ak}
\thanks{corresponding author; e-mail: andrej.bobak@upjs.sk}
\affiliation{Institute of Physics, Faculty of Science, P. J. \v{S}af\'arik University, Park Angelinum 9, 041 54 Ko\v{s}ice, Slovakia}
\author{T. Lu\v{c}ivjansk\'y}
\affiliation{Institute of Physics, Faculty of Science, P. J. \v{S}af\'arik University, Park Angelinum 9, 041 54 Ko\v{s}ice, Slovakia}
\affiliation{Fakult\"at f\"ur Physik, Universit\"at Duisburg-Essen, D-47048 Duisburg, Germany}
\author{M. \v{Z}ukovi\v{c}}
\affiliation{Institute of Physics, Faculty of Science, P. J. \v{S}af\'arik University, Park Angelinum 9, 041 54 Ko\v{s}ice, Slovakia}
\author{M. Borovsk\'y}
\affiliation{Institute of Physics, Faculty of Science, P. J. \v{S}af\'arik University, Park Angelinum 9, 041 54 Ko\v{s}ice, Slovakia}
\author{T. Balcerzak}
\affiliation{Department of Solid State Physics, University of \L\'{o}d\'{z}, Pomorska 149/153, 90-236 \L\'{o}d\'{z}, Poland}

\begin{abstract}
We study effects of the next-next-nearest-neighbour antiferromagnetic ($J_3 < 0$) interaction on critical properties (or phase diagram) of the frustrated spin-$\frac{1}{2}$ $J_1-J_2-J_3$ Ising antiferromagnet on the honeycomb lattice by using the effective-field theory with correlations. Beside the ground-state energy, we find that there is a region of $J_3 < 0$ in which the frustrated honeycomb lattice antiferromagnet exhibits a tricritical point, at which the phase transition changes from the second order to the first one on the line between N\'eel antiferromagnetic and paramagnetic phases.  
\end{abstract}

\pacs{05.50.+q, 05.70.-a, 75.10.Hk }

\maketitle

%
\section{Introduction}
Since a honeycomb lattice antiferromagnet with only nearest-neighbour (nn) interactions ($J_1 < 0$) is considered as a bipartite lattice, the ground state exhibits long-range ordering. However, the system is rather fragile against the onset of frustrating interactions. In recent years, therefore, it has become of great interest to investigate the corresponding model where the nn bonds are augmented by frustrating next-nearest-neighbour (nnn) bonds with the strength $J_2 < 0$, possibly also in conjunction with next-next-nearest-neighbour (nnnn) bonds of the strength $J_3 < 0$. An interest in the honeycomb lattice is also promoted from the recent experimental activity \cite{ref1} and from graphen-related issues \cite{ref2}. Due to these reasons, recently there has been a huge theoretical interest in frustrated spin models on the honeycomb lattice, in which frustration is incorporated by nnn interactions and maybe also nnnn interactions \cite{ref3}.  
In this paper we utilize the effective-field theory with correlations (EFT) as in our earlier work for the simpler $J_1-J_2$ model \cite{ref4}. Therefore, it will be interesting to study effects of frustration on the phase diagram of this bipartite lattice without making the restriction $J_3 = 0$. 
\section{Formulation}
We consider the frustrated honeycomb Ising antiferromagnet (AF) with nn $(J_1 < 0)$, nnn $(J_2 < 0)$, and nnnn $(J_3 < 0)$ interactions. Then the Hamiltonian for the resulting spin-$\frac{1}{2}$ $J_1-J_2-J_3$ Ising AF on the honeycomb lattice is given by 
\begin{equation}
\label{ham}
H = -J_1\sum_{\langle i,j\rangle}s_i s_j - J_2 \sum_{\langle i,i_{2}\rangle}s_i s_{i_2} - J_3 \sum_{\langle i,i_{3}\rangle}s_i s_{i_3},
\end{equation}
where index $i$ runs over all honeycomb lattice sites, and indices $j, i_2$, and $i_3$ run over all nn, nnn, and nnnn sites to $i,$ respectively, counting each bond once and once only, and $s_i = \pm 1$. The lattice and interactions are illustrated in Fig. 1.
\begin{figure}[h!]
\includegraphics[width=0.9\columnwidth]{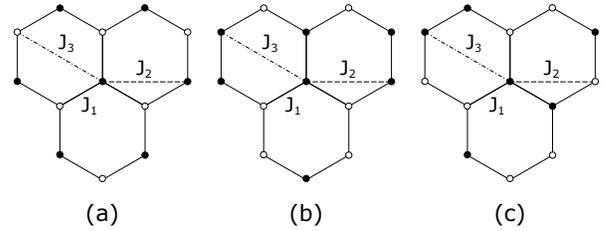}
\caption{The $J_1-J_2-J_3$ Ising model on the honeycomb lattice, showing in (a) the N\'eel state, where two sublattices are marked by black and white circles. The anti-N\'eel states are shown in (b) and (c) (see text).}
\label{fig1}
\end{figure}
We discuss first the ground state of this model. The AF phases consist of the N\'eel phase (N) (Fig. 1(a)) and two anti-N\'eel phases (aN) described either by alternate single ferromagnetic columns of antiparallel spins or alternate pairs of columns consisting of AF coupled spins shown in Fig. 1(b) and Fig. 1(c) (see also Fig. 17 of \cite{ref5}). The ground-state energy per site for the AF N\'eel phase is  
\begin{equation}
\label{AF}
\frac{E_{N}}{N} = -\frac{3}{2} (|J_1| + 2 J_2 - J_3).
\end{equation}   
In this case each site has its three nn bonds on the other sublattice, six nnn on its own sublattice, and three nnnn on the other sublattice. Thus only the nnn interactions act to frustrate the antiferromagnetism. On the other hand, the aN ground state for $J_3 =0$ is twofold degenerate and two states, labeled (b) and (c) in Fig. 1, have exactly the same ground-state energy. However, adding the nnnn AF interactions (or $J_3 < 0$), only the (c) state has lower energy given by 
\begin{equation}
\label{aN}
\frac{E_{aN}}{N} = -\frac{1}{2}(|J_1| - 2 J_2 + J_3).
\end{equation}
Now, pairwise equating the ground-state energies of the different AF phases, one can find that the first-order transition between the N and aN phases is determined by
\begin{equation}
\label{trans}
R = -\frac{1}{4} + \frac{J_3}{n|J_1|},
\end{equation}    
where $R = J_2/|J_1|$ is the frustration parameter and $n =~2$. On the other hand, we have found by using Monte Carlo simulations that the ground state of the aN phase is degenerate and has the energy 
\begin{equation}
\label{aNMC}
\frac{E_{aN}}{N} = -\frac{1}{2}(|J_1| - 2 J_2 - J_3).
\end{equation} 
Therefore, the ground-state energy per site in this case is less than that given by relation (\ref{aN}). Comparing this ground-state energy of the aN phase with (\ref{AF}), we obtain that the N and aN phases have the same energy only on the line given by Eq. (\ref{trans}) with $n = 4$.   

Let us consider the EFT (for a review see, e.g., Ref. \cite{ref6}) based on a single-site cluster containing only one spin on a site $i$ and a sublattice $A$ which interacts with other nn, nnn, and nnnn spins from the neighbourhood. In this approach, applying the differential operator technique, and using the van der Waerden identity for the two-state Ising spin system, one finds for the AF cluster on the honeycomb lattice the exact relation
\begin{equation}
\begin{array}{r l}
\langle s_i^A \rangle =& \Big\langle \prod\limits_{i_1=1}^{3}(A_1 + B_1 s_{i_1}^B)\prod\limits_{i_2=1}^{6}(A_2 + B_2 s_{i_2}^A) \\
& \times \prod\limits_{i_3=1}^{3}(A_3 + B_3 s_{i_3}^B) \Big\rangle \tanh(\beta x)\Big|_{x=0},
\end{array}
\label{exactmagA}
\end{equation} 
where $\langle \cdots \rangle$ denotes a thermal average, $s_{i_2}^A$ and  $s_{i_1}^B, s_{i_3}^B$ are spin variables on sublattices $A$ and $B$, respectively, $A_\nu = \cosh(J_\nu D_x)$, $B_\nu = \sinh(J_\nu D_x)$ $(\nu = 1, 2, 3)$, $D_x = \partial/\partial x$ is the differential operator, and $\beta = 1/{k_BT}$. Now, assuming the statistical independence of lattice sites, Eq. (\ref{exactmagA}) reduces to      
\begin{equation}
\begin{array}{r l}
m_A = &(A_1 + B_1 m_B)^3(A_2 + B_2 m_A)^6 \\ 
 &\times (A_3 + B_3 m_B)^3 \tanh(\beta x)\Big|_{x=0}, 
\end{array}
\label{mag_A}
\end{equation}
where $m_\alpha = \langle s_g^\alpha \rangle $ ($\alpha = A, B$) are the sublattice magnetizations per site. It should be noted here that this approximation is quite superior to the standard mean-field theory since even though it neglects correlations between different spins but takes the single-site kinematic relations exactly into account through the van der Waerden identity. On the other hand, the standard mean-field theory neglects all correlations. 

At this place, in order to solve the problem generally, we need to evaluate the sublattice magnetization $m_B$. It can be derived in the same way as $m_A$ by the use of the selected spin $s_j$ on the $B$ sublattice. However, at zero magnetic field we have $m_S \equiv m_A = -m_B$ and the equation for $m_B$ is the same as Eq. (\ref{mag_A}). Therefore, in what follows we use only Eq. (\ref{mag_A}), which in this case takes the final form 
\begin{equation}
\label{mag_S}
m_S = \sum_{{n}=0}^{5} K_{2n+1} m_S^{2n+1}. 
\end{equation}
The coefficients $K_{2n+1}$, which depend on $T, R$, and $J_3$, can be easily calculated within the symbolic programming by using the mathematical relation $\exp(\lambda D_x) f(x) = f(x+\lambda)$. 

In order to determine the phase diagram of the AF $J_1-J_2-J_3$ model, we should solve Eq. (\ref{mag_S}) for a given value of the frustration parameter $R$ and the interaction $J_3$, and look for the temperature at which the magnetization (order parameter) $m_S$ goes to zero. However, for some values of $R$ and $J_3$, the order parameter goes to zero discontinuously, i.~e., the transition becomes first order. To analyze first-order transitions, one needs to calculate the free energy for the N and paramagnetic (P) phases and to find a point of intersection. Because the expression for the free energy in this effective-field theory does not exist, it will be extrapolated with the help of the relation for the equilibrium value of the order parameter (\ref{mag_S}) as follows \cite{ref7}:
\begin{equation}
\label{freeenergy}
F(T, R, J_3, m) = F_0(T, R, J_3) + \frac{1}{2}\Big(1-\sum_{{n}=0}^{5} \frac{K_{2n+1}}{n+1}{m^{2n}}\Big)m^2, 
\end{equation}
where $F_0(T,R, J_3)$ is the free energy of the disordered (paramagnetic) phase and $m$ is the order parameter which takes the value $m_S$ at thermodynamic equilibrium. We note that relation (\ref{freeenergy}) corresponds to a Landau free energy expansion in the order parameter truncated at the $m^{12}$ term. Then a critical temperature and a tricritical point, at which the phase transition changes from second order to first order, are determined by the following conditions \cite{ref7}: (i) the second-order transition line when $1-K_1 =~0$ and $K_3 < 0,$ and (ii) the tricritical point (TCP) when $1-K_1 = 0$, $K_3 = 0,$ if $K_5 < 0$. However, the first-order phase transition line is evaluated by solving simultaneously two transcendental equations, namely the equilibrium condition $[\partial F(T, R, J_3, m)/\partial m]_{m=m_S} = 0$ and the equation $F(T, R, J_3, m) = F_0(T, R, J_3)$ that corresponds to the point of intersection of the free energies for the N and P phases. 
\section{Results and discussion}
Now, by using the general formulation given in the previous section, let us examine the phase diagram of the system. 
\begin{figure}[h!]
\includegraphics[width=1.0\columnwidth,height=5.0cm]{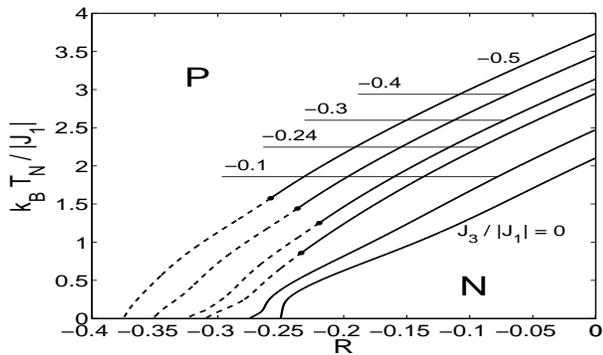}
\caption{Phase diagram in the $R-T$ plane for the $J_1-J_2-J_3$ model, when the interaction $J_3/|J_1|$ is changed. Solid and dashed lines indicate second- and first-order transitions, respectively, while the black circles denote the position of a TCP. N and P are the N\'eel and paramagnetic phases, respectively.}
\label{fig2}
\end{figure}
In Fig. 2, the critical temperature $k_BT_N/|J_1|$ versus $R$ is shown for selected values of $J_3/|J_1|$. The solid lines indicate the second-order phase transitions, while the dashed lines represent the first-order phase transitions. The black circles denote the position of TCPs at which the phase transitions change from second to first order. The most important feature in Fig. 2 is that the $J_1-J_2$ model on the honeycomb lattice exhibits only the second-order phase transition which vanishes at $R = -1/4$, in agreement with the ground state discussed above. However, if the $J_3$ interaction is different from zero and gradually decreases from zero to a larger negative value, the tricritical point appears in the system for $J_3/|J_1| = - 0.2396$ (see, e.g. the curve labeled by -0.24). It is also noteworthy that all transition temperatures $T_N$ between the N and P phases as function of the frustration parameter $R$ approach zero at the values of $R$ determined by Eq. (\ref{trans}) with $n = 4$. This indicates that the ground-state energy of the aN phase for the $J_1-J_2-J_3$ Ising model on the honeycomb lattice is indeed given by Eq. (\ref{aNMC}).  
\begin{figure}[h!]
\includegraphics[width=1.0\columnwidth,height=5.0cm]{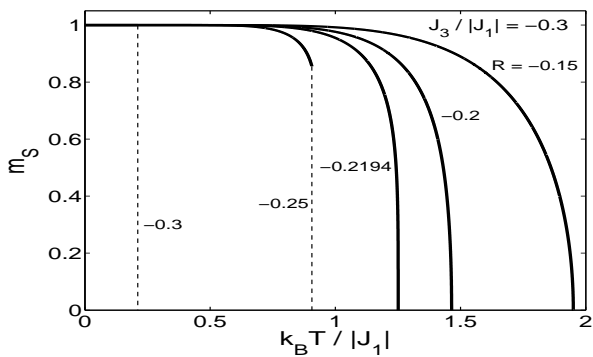}
\caption{Temperature dependence of the order parameter of $m_S$ for the $J_1-J_2-J_3$ Ising model with $J_3/|J_1|=-0.3$, when the frustration parameter $R$ is changed. The dashed lines indicate the first-order transitions.}
\label{fig3}
\end{figure}

In order to confirm the prediction of the first- and second-order phase transitions, let us examine temperature dependencies of the order parameter $m_S$ for the system with $J_3/|J_1| = -0.3$, when the value of $R$ is changed. As can be seen in Fig. 3, the order parameter $m_S$ falls smoothly to zero when temperature increases from zero to $k_BT_N/|J_1|$, characterizing a second-order phase transition. Similarly, $m_S$ also reduces to zero continuously at the TCP (see curve labeled $-0.2194$). On the other hand, below the TCP, the stable solution of $m_S$ becomes discontinuous at the first-order phase transition and this discontinuity increases with $R$ going to $-0.3250$. The curves for $R = -0.25$ and $R= -0.3$ are examples of such behavior, where the first-order transition is indicated by a vertical dashed line. 

\section{Conclusions}
We have studied the phase diagram in the $(R, T)$ plane of the frustrated $J_1-J_2-J_3$ Ising model with spin-$\frac{1}{2}$ on a honeycomb lattice using the EFT based on the single-spin cluster. We have determined that in the ground-state two ordered phases, namely the N and aN states coexist only on the line given by Eq. (\ref{trans}) with $n = 4$. However, for the aN phase we have not found a long-range order at $T \neq 0$ K due to the degeneracy of the ground state. This behaviour has been also confirmed by our preliminary Monte Carlo calculations. On the other hand, the present EFT predicts the TCP in the phase diagram between the N and P phases due to the $J_3$ interaction. Of course, this is the effective-field result, therefore, further Monte Carlo simulations or more reliable calculations for this frustrated $J_1-J_2-J_3$ Ising model would be desirable.    
\section{Acknowledgement}
This work was supported by the Grant VEGA 1/0331/15. 

\end{document}